\documentclass[aps,preprint]{revtex4}
\usepackage{graphicx}% Include figure files
\makeatletter
  \newcommand{\figcaption}{\def\@captype{figure}\caption}
  \newcommand{\tabcaption}{\def\@captype{table}\caption}
\makeatother

\begin{document}
%\draft

 \title {\bf Pairing correlations and resonant states\\ in the relativistic mean field theory}

\author{ N. Sandulescu$^{a),b),c)}$, L.S. Geng$^{c),d)}$, H. Toki $^{c)}$, and G. Hillhouse
$^{c),e)}$}

\vspace {03mm}

\address{
{\it a) Royal Institute of Technology, SCFAB, SE-10691, Stockholm, Sweden}\\
{\it b) Institute for Physics and Nuclear Engineering, P.O. Box
MG-6, 76900
Bucharest, Romania}\\
{\it c) Research Center for Nuclear Physics (RCNP),Osaka
University, 10-1, Mihogaoka, Ibaraki, Osaka 567-0047, Japan}\\
{\it d)School of Physics, Peking University, Beijing 100871, P. R.
China}\\
 {\it e) University of the Western Cape, Bellville 7530,
South Africa}\\ }

%\date{today}

\begin{abstract}
 We present a simple scheme for taking into account the resonant
 continuum coupling in the Relativistic Mean Field- BCS (RMF-BCS)
 calculations. In this scheme, applied before in  non-relativistic
 calculations, the effect of the resonant continuum on pairing
 correlations is introduced through the scattering wave functions
 located in the region of the resonant states. These states are
 found by solving the relativistic mean field equations with
 scattering-type boundary conditions for continuum spectrum.
 The calculations are done for the neutron-rich Zr isotopes.
 It is shown that the sudden increase of the neutron radii close
 to the neutron drip line, the so-called giant halo, is determined
 by a few resonant states close to the continuum threshold.
\end{abstract}

\maketitle

\section {Introduction}

 As recognized long time ago \cite{BMP}, the basic features of the superfluidity are
 the same in atomic nuclei and  infinite Fermi systems. Yet, in atomic nuclei the
 pairing correlations have many features related to the finite size of
 the system. The way how the finite size affects the pairing correlations depends on
 the position of the  chemical potential. If the chemical potential is deeply bound,
 like in stable and heavy nuclei, the finite size influences the  pairing
 correlations mainly  through the shell structure induced by the spin-orbit
 interaction. The situation becomes more complex in nuclei close to the drip lines,
 where the chemical potential approaches the continuum threshold.
 In this case the inhomogeneity of the pairing field produces strong
 mixing between the bound and  the continuum part of the single-particle spectrum.
 Due to this mixing the quasi-particle spectrum becomes dominated by resonant
 quasi-particle states, which originate both from single-particle
 resonances and deep hole states \cite{Belyaev,Fayans,Bulgac,Grasso1}.

 The continuum effects on pairing correlations is commonly taken into account in
 the Hartree-Fock-Bogoliubov (HFB) \cite{RS} or Relativistic-Hartree-Bogoliubov (RHB)
 \cite{Ring} approach. In most of these calculations the continuum is replaced
 by a set of positive energy states determined by solving the HFB or RHB
 equations in coordinate space and with box boundary conditions  \cite{Doba,Meng1}.
 Due to this fact the genuine continuum  effects, as the widths of the quasi-particle
 states, are not accounted for in these type of calculations.

 Recently the HFB equations were also solved with exact boundary conditions for the
 continuum spectrum, both for a zero range \cite{Grasso1} and a finite range pairing force
 \cite{Grasso2}. It was thus shown that close to the drip lines the discretisation
 of the continuum generally overestimate the pairing correlations. A similar
 conclusion was obtained earlier in a simpler BCS approach, in which the resonant
 part of the continuum was studied \cite{Sandulescu1,Sandulescu2}. Comparing these
 BCS results with the exact HFB solutions \cite{Grasso1} one finds that the effect
 of the continuum on pairing correlations is  given mainly by a few resonant
 states close to the continuum threshold.

 For the relativistic models an exact solution of the continuum spectrum
 is not available yet, neither for RHB nor for  RMF-BCS approach.
 A comparison between RHB and RMF-BCS calculations, performed by 
 using box boundary conditions, is discussed in Ref.\cite{Estal}. 
 This comparison indicates also the special role played by the resonant 
 states, which in these calculations are approximated by positive energy states.
 This approximation works well only if the positive energy states
 correspond to very narrow resonances. Moreover, since a discrete
 representation of the continuum does not provide a direct measure
 of  the width of the resonant states, the selection of the relevant
 positive energy states is ambiguous if the resonant states close
 to the continuum threshold are not very narrow.

 The scope of this paper is to show how the resonant continuum can be
 treated accurately in the RMF-BCS approach. The single-particle states belonging
 to the resonant part of the continuum spectrum will be calculated by solving the
 RMF equations with scattering-type boundary conditions. Then the resonant
 continuum will be handled in the BCS equations in a similar way as in the
 non-relativistic HF-BCS calculations \cite{Sandulescu2}.
 This approach is applied for the case of Zr isotopes, for which earlier
 calculations predict a very large neutron skin close to the neutron drip line.
 It is shown that the sudden increase of the nuclear radii in these isotopes
 is essentially determined by a few single-particle resonant states close to the 
 continuum threshold.

 The article is organized as follows. In Section 2 we discuss shortly
 the scattering type solutions of the relativistic mean field equations
 and we introduce the resonant-BCS equations \cite{Sandulescu2}.
 Then in Section 3 we present the results of the calculations for Zr
 isotopes. Section 4 contains the summary of the paper.

\section {Resonant states in the RMF-BCS approach}

\subsection{Continuum-RMF solutions}

In the relativistic mean field approach the nuclear interaction is
usually described by the exchange of three mesons: the scalar
meson $\sigma$, which mediates the medium-range attraction between
the nucleons, the vector meson $\omega$, which mediates the
short-range repulsion, and the isovector-vector meson
$\vec{\rho}$, which provides the isospin dependence of the nuclear
force. The equations of motion are commonly derived from the
effective Lagrangian density \cite{rheinhard,Ring}:
\begin{eqnarray}
       {\cal L}& = &{\bar\psi} [\imath \gamma^{\mu}\partial_{\mu}
                  - M]\psi\nonumber\\
                  &&+ \frac{1}{2}\, \partial_{\mu}\sigma\partial^{\mu}\sigma
                - \frac{1}{2}m_{\sigma}^{2}\sigma^2- \frac{1}{3}g_{2}\sigma
                  ^{3} - \frac{1}{4}g_{3}\sigma^{4} -g_{\sigma}
                 {\bar\psi}  \sigma  \psi\nonumber\\
                &&-\frac{1}{4}H_{\mu \nu}H^{\mu \nu} + \frac{1}{2}m_{\omega}
                   ^{2}\omega_{\mu}\omega^{\mu} + \frac{1}{4} c_{3}
                  (\omega_{\mu} \omega^{\mu})^{2}
                   - g_{\omega}{\bar\psi} \gamma^{\mu}\psi
                  \omega_{\mu}\nonumber\\
               &&-\frac{1}{4}G_{\mu \nu}^{a}G^{a\mu \nu}
                  + \frac{1}{2}m_{\rho}
                   ^{2}\rho_{\mu}^{a}\rho^{a\mu}
                   - g_{\rho}{\bar\psi} \gamma_{\mu}\tau^{a}\psi
                  \rho^{\mu a}\nonumber\nonumber\\
                &&-\frac{1}{4}F_{\mu \nu}F^{\mu \nu}
                  - e{\bar\psi} \gamma_{\mu} \frac{(1-\tau_{3})}
                  {2} A^{\mu} \psi\,\,,%\nonumber\
\end{eqnarray}
where a non-linear self-coupling is considered both for $\sigma$ and $\omega$ mesons.
The vector fields H, G and F are given by
\begin{eqnarray}
                 H_{\mu \nu} &=& \partial_{\mu} \omega_{\nu} -
                       \partial_{\nu} \omega_{\mu}\nonumber\\
                 G_{\mu \nu}^{a} &=& \partial_{\mu} \rho_{\nu}^{a} -
                       \partial_{\nu} \rho_{\mu}^{a}
                     -2 g_{\rho}\,\epsilon^{abc} \rho_{\mu}^{b}
                    \rho_{\nu}^{c} \nonumber\\
                  F_{\mu \nu} &=& \partial_{\mu} A_{\nu} -
                       \partial_{\nu} A_{\mu}\,\,\nonumber\
\end{eqnarray}
The nucleons are described by the Dirac spinor field $\psi$, which in the case
of spherical symmetry can be written as:
\begin{equation}
\psi ={1 \over r} \,\, \left({i \,\,\, G \,\,\,
 {\mathcal Y}_{j l m}
\atop{F \, {\sigma} \cdot \hat{r}\,
\, {\mathcal Y}_{j l m}}} \right)\,\,,
\end{equation}
where ${\mathcal Y}_{jlm}$ denotes the spinor spherical harmonics
while $G$ and $F$ are the radial wave
functions for the upper and lower components, respectively.
They satisfy the radial equations:
\begin{equation}
 \frac{dG}{dr} +\frac{\kappa}{r}G - (M+E+V_{s}-V_{v})F = 0
\end{equation}
\begin{equation}
 -\frac{dF}{dr} +\frac{\kappa}{r}F + (M-E+V_{s}+V_{v})G = 0
\end{equation}
where $V_{s}$ and $V_{v}$ are the scalar and the vector mean fields
and $\kappa$ is given by:
\begin{equation}
\kappa = \left\{  \begin{array}{ll}
                    -(l+1) & \mbox{ if $j=l+1/2$}\\
                     +l    & \mbox{ if $j=l-1/2$}
                    \end{array}
          \right .
\end{equation}
At large distances, where both the scalar and the vector mean fields are zero,
the radial equations can be written in the form:
\begin{equation}
 \frac{d^2G}{dr^2} +(\alpha^2- \frac{\kappa(\kappa+1)}{r^2})G  =  0
\end{equation}
\begin{equation}
 F  =  \frac{1}{E+M}(\frac{dG}{dr}+\frac{\kappa}{r}G) ,
\end{equation}
where $\alpha^2=E^2-M^2$. These equations are suited for fixing
the scattering-type boundary conditions for the continuum spectrum. They are
given by:
\begin{equation}
 G =  C \alpha r [cos(\delta) j_l(\alpha r)-sin(\delta) n_l(\alpha r)]
\end{equation}
\begin{equation}
 F =  \frac{C \alpha^2r}{E+M}
[cos(\delta) j_{l-1}(\alpha r)-sin(\delta) n_{l-1}(\alpha r)] ,
\end{equation}
where $j_{l}$ and $n_{l}$ are the Bessel and  Neumann functions and
$\delta$ is the phase shift associated to the relativistic mean field.
The constant $C$ is fixed by the normalisation condition of the
scattering wave functions and the phase shift $\delta$ is calculated
from the matching conditions.
In the vicinity of an isolated resonance the derivative of the phase
shift has a Breit-Breit form, i.e.,
\begin{equation}
\frac{d\delta(E)}{dE} = \frac{\Gamma/2}{(E_r-E)^2+\Gamma^2/4}
\end{equation}
from which one estimates the energy and the width of the resonance.
In the vicinity of a resonance the radial wave functions of the
scattering states have a large localisation inside the nucleus.
Close to a resonance the energy dependence of both components of the
Dirac wave functions can be factorized approximatively by  
a unique energy dependent function \cite{Migdal}. As in the non-relativistic
case \cite{Unger}, this energy dependent
factor is the square root of the Breit-Wigner function written above, or, 
equivalently,  the square root of the derivative of the phase shift.
 Using this property all the matrix elements of a two-body interaction 
between these scattering states can be expressed in term of a unique 
matrix element, i.e. the one corresponding to the scattering state with
energy equal to the energy of the resonance. This property is employed below
for the treatment of the resonant continuum in the BCS equations.

\subsection{Resonant states in the BCS approach}

 Since the meson exchange forces are not describing properly the pairing
 correlations in nuclei, the relativistic mean field is combined usually
 with non-relativistic pairing models. Here we use for the pairing
 treatment the BCS approach. The extension of the BCS equations for taking
 into account the resonant continuum was proposed
 in Refs. \cite{Sandulescu1,Sandulescu2}. For the case of a general pairing
 interaction these equations, referred below as the resonant-BCS (rBCS) equations,
 reads \cite{Sandulescu2}:

\begin{equation}\label{eq:gapr1}
\Delta_i = \sum_{j}V_{i\overline{i}j\overline{j}} u_j v_j +
\sum_\nu
V_{i\overline{i},\nu\epsilon_\nu\overline{\nu\epsilon_\nu}}
\int_{I_\nu} g_{\nu}(\epsilon)
u_\nu(\epsilon) v_\nu(\epsilon) d\epsilon~,
\end{equation}
\begin{equation}\label{eq:gapr}
\Delta_\nu \equiv
\sum_{j}
V_{\nu\epsilon_\nu\overline{\nu\epsilon_\nu},j\overline{j}} u_j v_j +
\sum_{\nu^\prime}
V_{\nu\epsilon_\nu\overline{\nu\epsilon_\nu},
\nu^\prime\epsilon_{\nu^\prime}
\overline{\nu^\prime\epsilon_{\nu^\prime}}}
\int_{I_{\nu^\prime}} g_{\nu^\prime}(\epsilon^\prime)
u_{\nu^\prime}(\epsilon^\prime) v_{\nu^\prime}(\epsilon^\prime)
d\epsilon^\prime~,
\end{equation}
\begin{eqnarray}
N = \sum_i v_i^2 + \sum_\nu \int_{I_\nu} g^c_{\nu}(\epsilon) v^2_\nu
(\epsilon) d\epsilon~.
\label{eq15}
\end{eqnarray}
Here $\Delta_i$ are the gaps for the bound states and
$\Delta_\nu$ are the averaged gaps for the resonant states.
The quantity $g^c_\nu(\epsilon) = \frac {2j_\nu +1}{\pi}
\frac{d\delta_\nu}{d\epsilon}$ is the total level density and
$\delta_\nu$ is the phase shift of angular momentum $(l_{\nu} j_{\nu})$.
The factor $g^c_\nu(\epsilon)$ takes into account the variation
of the localisation of scattering states in the energy region of
a resonance ( i.e., the width effect) and goes  to a delta function in the
limit of a very narrow width. The interaction matrix elements are  calculated with
the scattering wave functions at  resonance energies and normalised
inside the volume where the pairing interaction is active.
For more details see Ref.\cite{Sandulescu2}.

The rBCS equations written above are applied here with the single-particle
spectrum of the RMF equations. For the pairing interaction we use in the
next section a delta force, i.e. $V=V_0 \delta(\vec{r}_1-\vec{r}_2)$.
In this case the matrix elements of the pairing interaction are given by:
\begin{equation}
\left<(\tau_1 \bar{\tau_1})\,0^+\,|V|\,(\tau_2 \bar{\tau_2})\,0^+\right>  =
\frac{V_0}{8\pi}
\int\,dr \frac{1}{r^2}\,\left(G^\star_{\tau_1}\, G_{\tau_2}\,+\,
     F^\star_{\tau_1}\, F_{\tau_2}\right)^2
\end{equation}
For the resonant states these matrix elements are calculated as mentioned
above, i.e. using the radial wave functions evaluated at resonance energies
and normalised inside a finite volume.

The RMF and the rBCS equations are solved iteratively. At each iteration
the densities are modified through the occupation probabilities provided
by the rBCS, as in the non-relativistic HF-rBCS calculations \cite{Sandulescu2}.

\section {RMF-rBCS calculations for neutron-rich Zr isotopes}

 Zr isotopes were discussed
recently in connection to the so-called giant halo structure,
which these isotopes may develop close to the neutron drip line
\cite{Meng2}.
 In Ref. \cite{Meng2} these isotopes were calculated by solving
the RHB equations in coordinate representation and using box
boundary conditions. The mean field was described by the
parameter set NLSH \cite{Sharma} and for the  pairing
interaction was employed a density dependent delta interaction.
In the calculations were introduced all the positive energy states
up to 120 MeV.

In order to compare our calculations with the RHB predictions of
Ref.\cite{Meng2} we use for the mean field the same parameter set,
i.e. NLSH. The results are not much different even when 
we use other parameter sets, e.g. NL3 and TM1 \cite{ring3,sugahara}. 
The appropriate choice for the pairing interaction is more difficult
because the pairing correlations estimated with a zero range force
depend strongly on the energy cut-off, which is very different in
the two calculations. Thus in the RMF-rBCS approach we include
from all the continuum only a few resonant states close to zero
energy while in the RHB calculations the pairs are virtually
scattered in all the positive energy states up to the energy cut
off, i.e. 120 MeV. This energy cut off is here much larger than
the maximum quasi-particle energy calculated in RMF-rBCS, which
corresponds to the single-particle bound state $1s_{1/2}$. Due to
these facts we cannot compare meaningfully the results of the two
calculations if we use the same zero range force. The best we can
do is to choose in the RMF-rBCS calculations a pairing force which
provides in average pairing energies close to the RHB values, at
least for some isotopes. Following this procedure we chose in the
RMF-rBCS calculations a zero range pairing force given by
$V=V_0({\bf r_1}-{\bf r_2})$, with $V_0$=- 275 MeV. We found that
the strength of the force can be actually increased up to about
$V_0$=-350 MeV without affecting significantly the separation
energies and the nuclear radii shown below.
\\[\intextsep]
\begin{minipage}{\textwidth}
\centering
\includegraphics[scale=0.7]{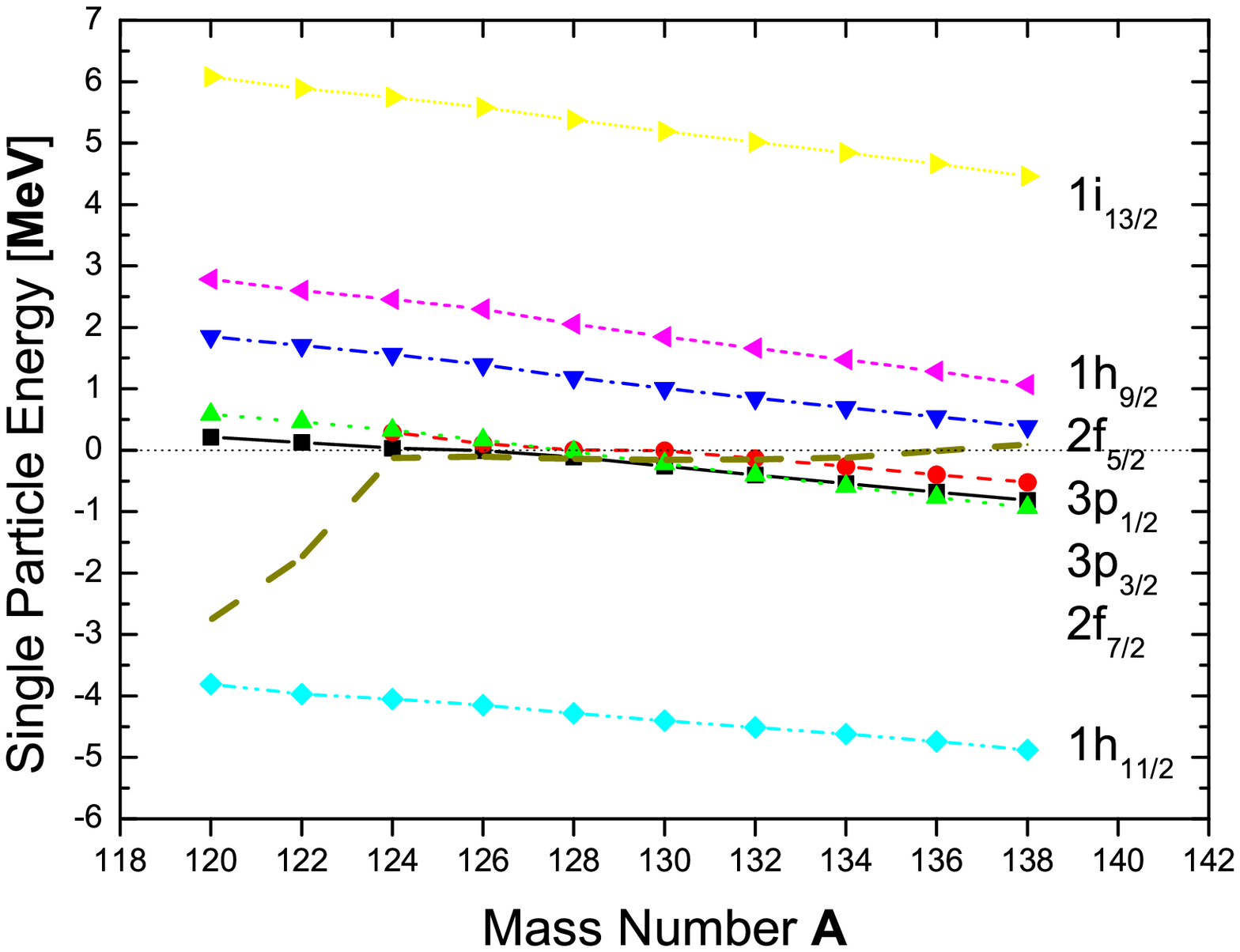}
\figcaption{\label{fig1.fig}The  energies of  bound and resonant
single-particle states close to the continuum threshold in Zr
isotopes. The Fermi energy is shown by the dashed line.}
\end{minipage}
\\[\intextsep]

First we analyse the behaviour of the single-particle states in
the vicinity of the continuum threshold. These states play the
major role in the formation of the neutron skin structure
discussed below. In Figure 1 are shown the neutron single-particle
levels for the $Zr$ isotopes closest to the neutron drip
line, i.e. from the mass number $A = 120$ up to $A=138$. The
dashed line represents the chemical potential,  which stays close
to zero from $A = 124$ to $A = 138$. The positive energies shown
here are the energies of the resonant states. As seen in Figure 1,
the states $2f_{5/2}$, $1h_{9/2}$ and $1i_{13/2}$ remain resonant
states for all the isotopes and their energies are changing with
the neutron number in the same way as the energy of the bound
state $1h_{11/2}$. The other three states, i.e. $2f_{7/2}$,
$3p_{3/2}$ and $3p_{1/2}$, are resonant states for $A < 126$ and
become loosely bound states for heavier isotopes. Their radial
wave functions for $A= 124 $ are shown in Figure 2. The figure
shows the upper components of the radial wave functions calculated
at the resonance energies, for which their localisation inside the
nucleus is the largest.
\\[\intextsep]
\begin{minipage}{\textwidth}
\centering
\includegraphics[scale=0.7]{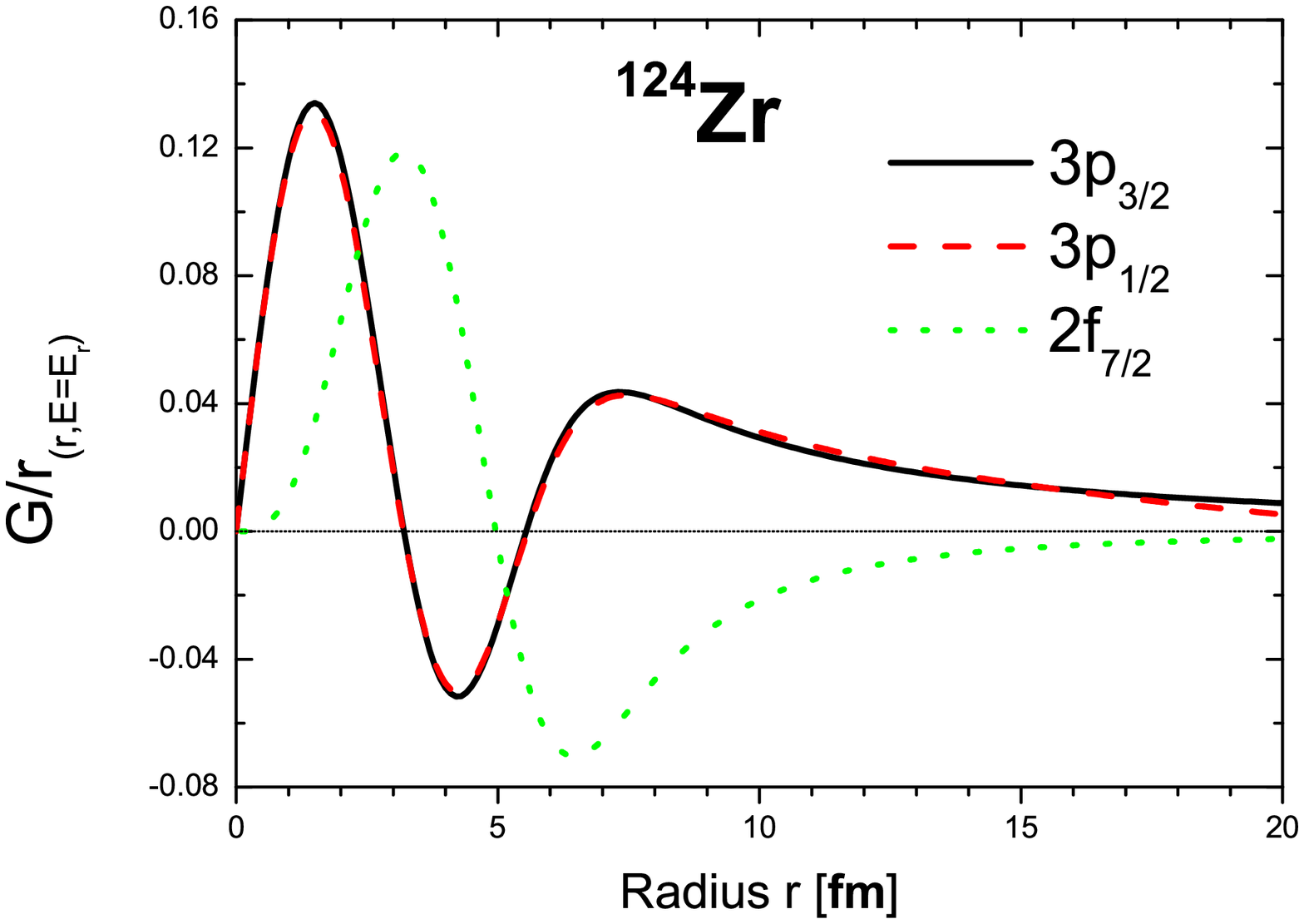}
\figcaption{\label{fig2.fig}The radial wave functions of the
resonant states $2f_{7/2}$, $3p_{3/2}$ and $3p_{1/2}$  in
$^{124}$Zr. The plot represents the upper components of the radial
wave functions calculated at the resonance energies.}
\end{minipage}
\begin{minipage}{\textwidth}
\centering
\includegraphics[scale=0.7]{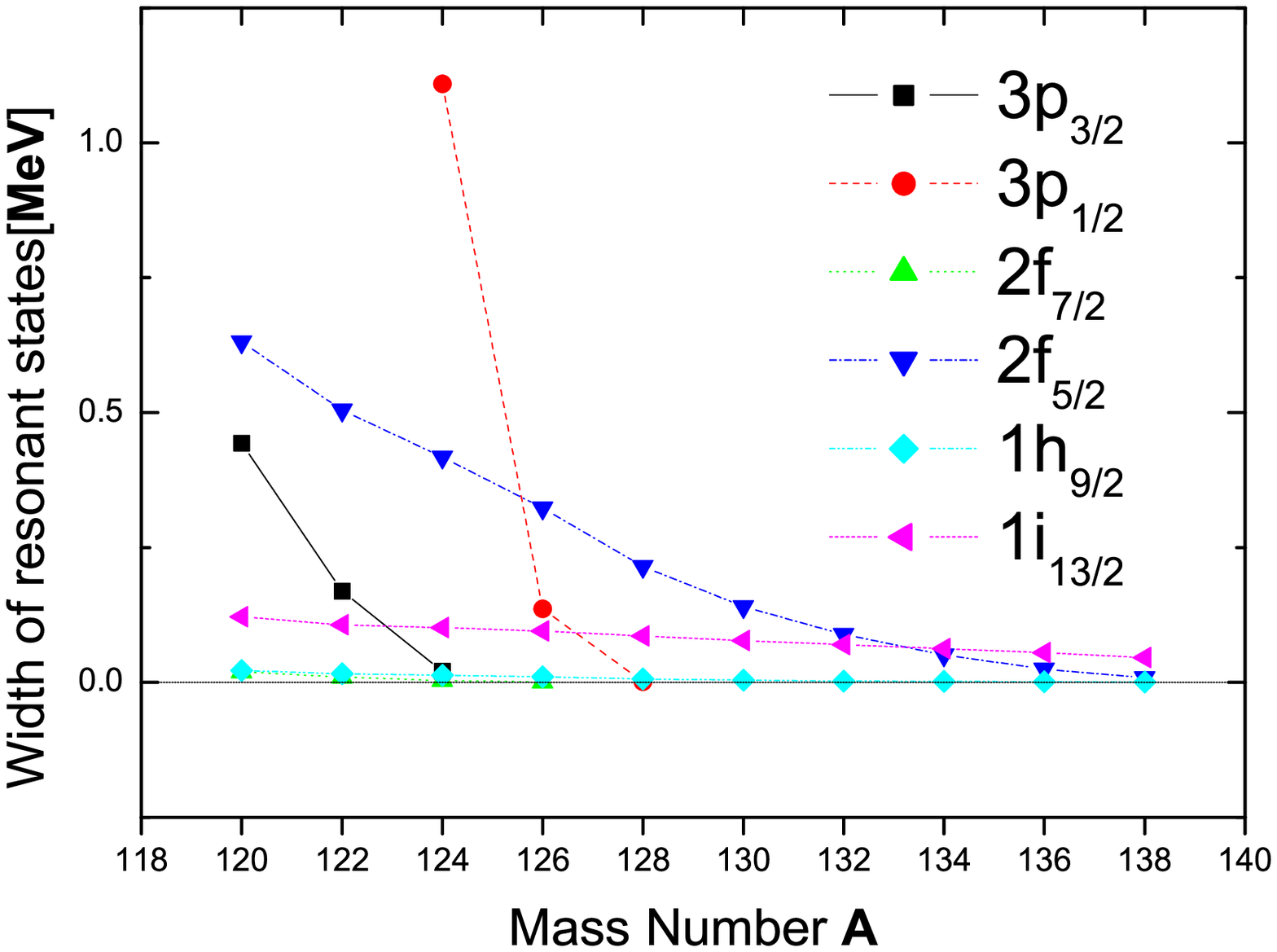}
\figcaption{\label{fig3.fig}The  width of the single-particle
resonant states shown in Figure 1.}
\end{minipage}
\\[\intextsep]

The widths of the resonant states are plotted in Figure 3. The
resonant states $1h_{9/2}$ and $2f_{5/2}$ have rather small widths
for all the isotopes.  On the other hand the widths of the
resonant states $3p_{1/2}$ and $3p_{3/2}$ are changing
dramatically with the neutron number. This is especially the case
for the resonant state $3p_{1/2}$. Considering that these  states
with low (lj) values have a major contribution to the formation of
the neutron skin, their wave functions should be calculated
accurately, both for positive and negative energies. The resonant
states with high (lj) values give instead the dominant
contribution to the pairing correlations. Since these resonant
states have a small width they could be eventually treated like
quasi-bound states in the pairing calculations.

Next we examine the two- neutron separation energies  $S_{2n}$,
i.e.,
\begin{equation}
S_{2n}(Z,N)=B(Z,N)-B(Z,N-2)
\end{equation}
Their values are shown in Figure 4. The empirical values correspond to
Ref.\cite{Audi} while the RHB results are the one given in Ref.\cite{Meng2}.
One can see that RMF-rBCS gives practically the same results as the
RHB calulaculations. The two-neutron separation energies remain close to
zero all the way from A=124 to A=138, which in RMF would corresponds to
the filling of the group of states $2f_{7/2}$, $3p_{3/2}$ and $3p_{1/2}$.
\\[\intextsep]
\begin{minipage}{\textwidth}
\centering
\includegraphics[scale=0.7]{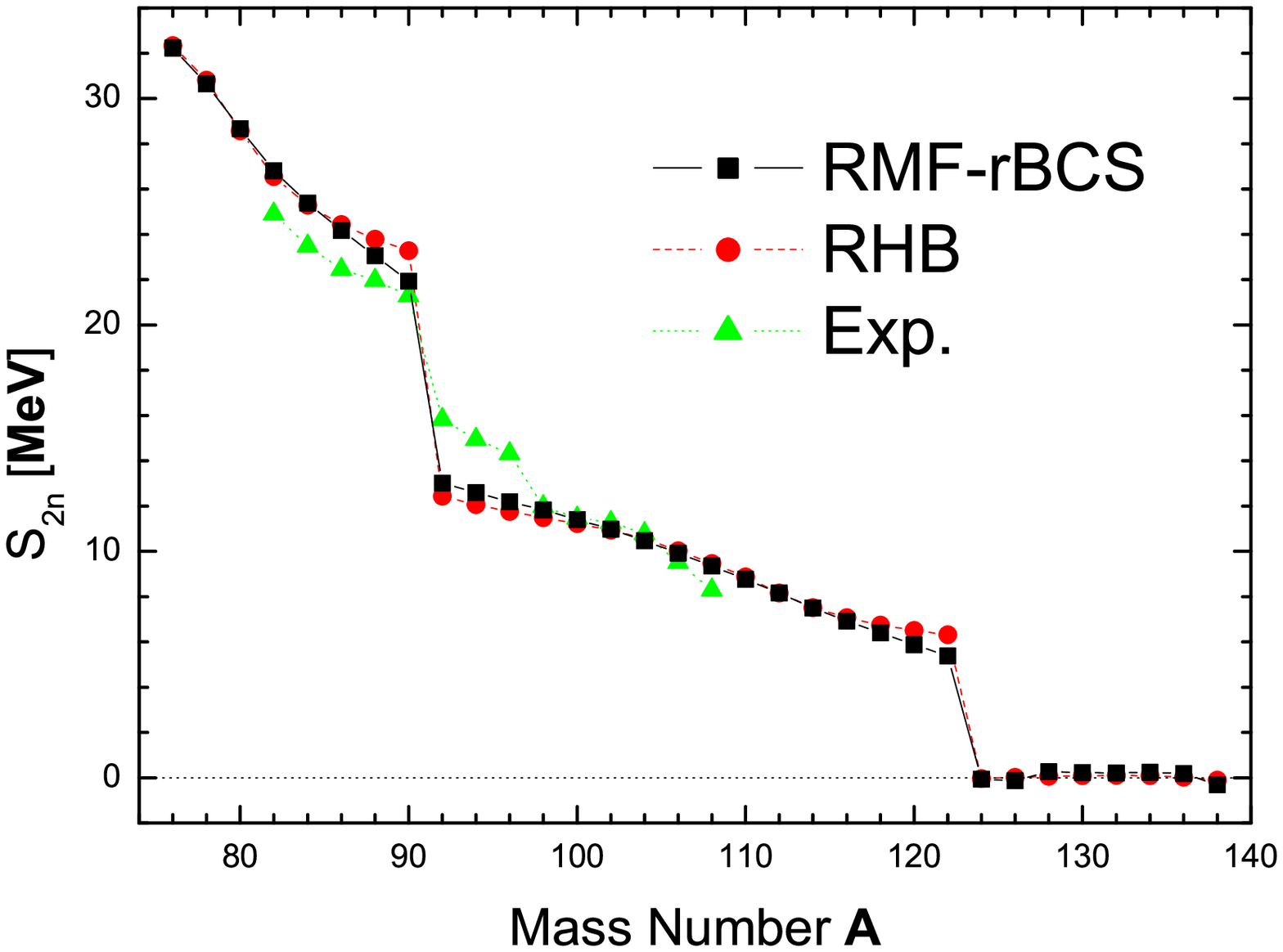}
\figcaption{\label{fig4.fig} Two-neutron separation energies of
even Zr isotopes as a function of the mass number A.}
\end{minipage}
\\[\intextsep]

In order to see the amount of the pairing correlations in these
isotopes we plotted in Figure 5 the pairing correlation energies,
i.e. the binding energies referred to the RMF values. The pairing
correlation energy curve shows a minimum for $A=136$, which
corresponds to the filling of the states $2f_{7/2}$, $3p_{3/2}$.
These states have almost the same energy and behave like a closed
major shell for the pairing correlation energy.

 As discussed in Ref.\cite{Grasso1,Sandulescu1}, the pairing correlations
 become usually stronger when the widths of resonant states is not taken
 into account, i.e. when the resonant states are considered as quasi bound
 states. This effect can be also seen in Figure 5, even if in the
 present case the differences in correlation energies are rather small.

  The most interesting phenomenon in these isotopes is the behaviour
 of the neutron radii, which are shown in Figure 6. First one notices
 that the RMF-rBCS results follow again very closely the RHB values.
 As shown in Figure 6, the neutron radii increase sharply  from $A=122$
 to $A=124$. This is mainly due to the filling of the states $3p_{3/2}$
 and $2f_{7/2}$, which in $A=124$ are very narrow resonances close to
 the continuum threshold. For $A=124$ the resonant  state $3p_{1/2}$ has
 almost the same energy as the resonant state $3p_{3/2}$ but its contribution
 to the radius is reduced because it has a rather large width.
\\[\intextsep]
\begin{minipage}{\textwidth}
\centering
\includegraphics[scale=0.7]{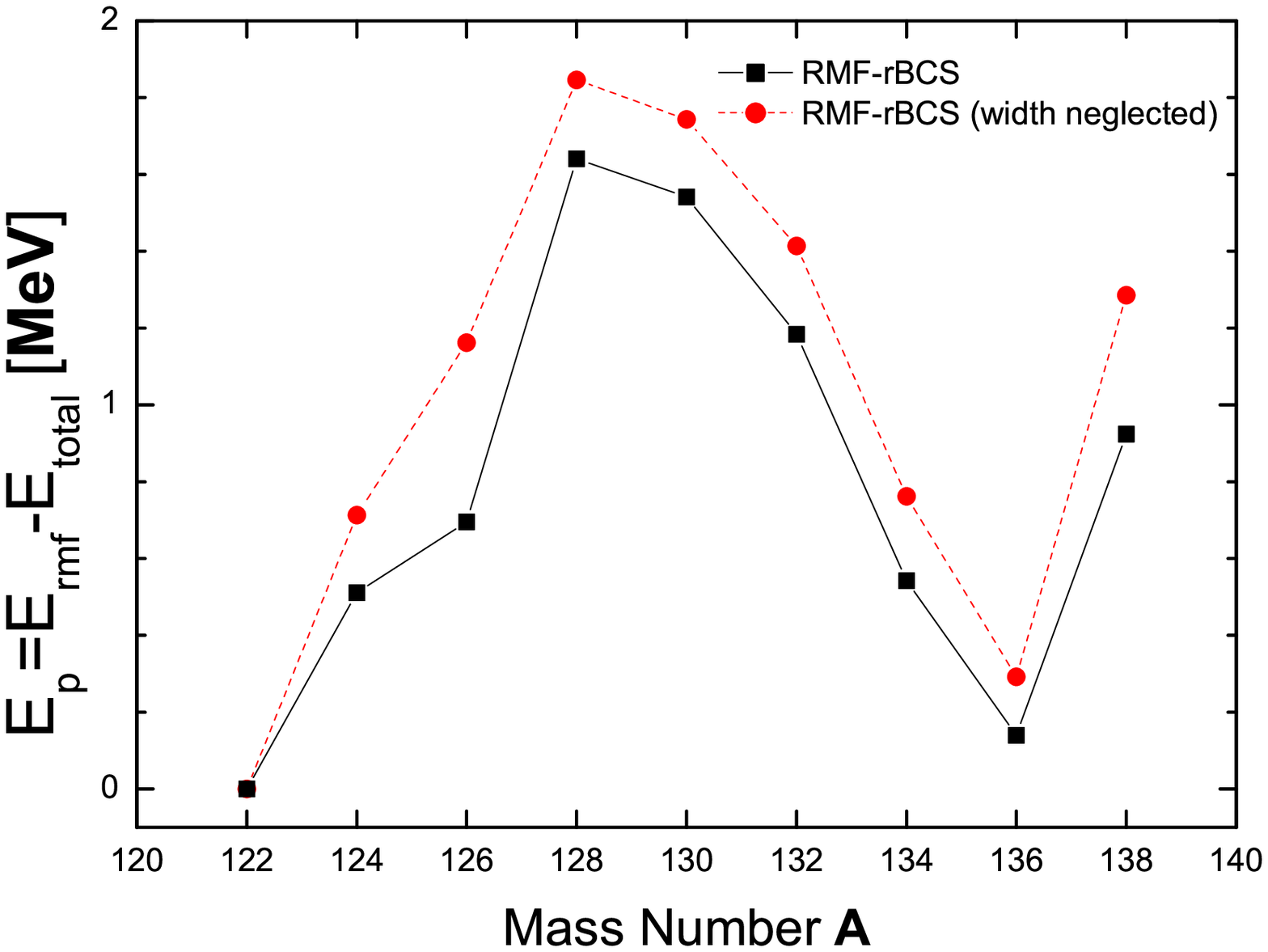}
\figcaption{\label{fig5.fig}Pairing correlation energies of even
Zr isotopes as a function of the mass number A.}
\end{minipage}
\\[\intextsep]

\begin{figure}
\centering
\includegraphics[scale=0.7]{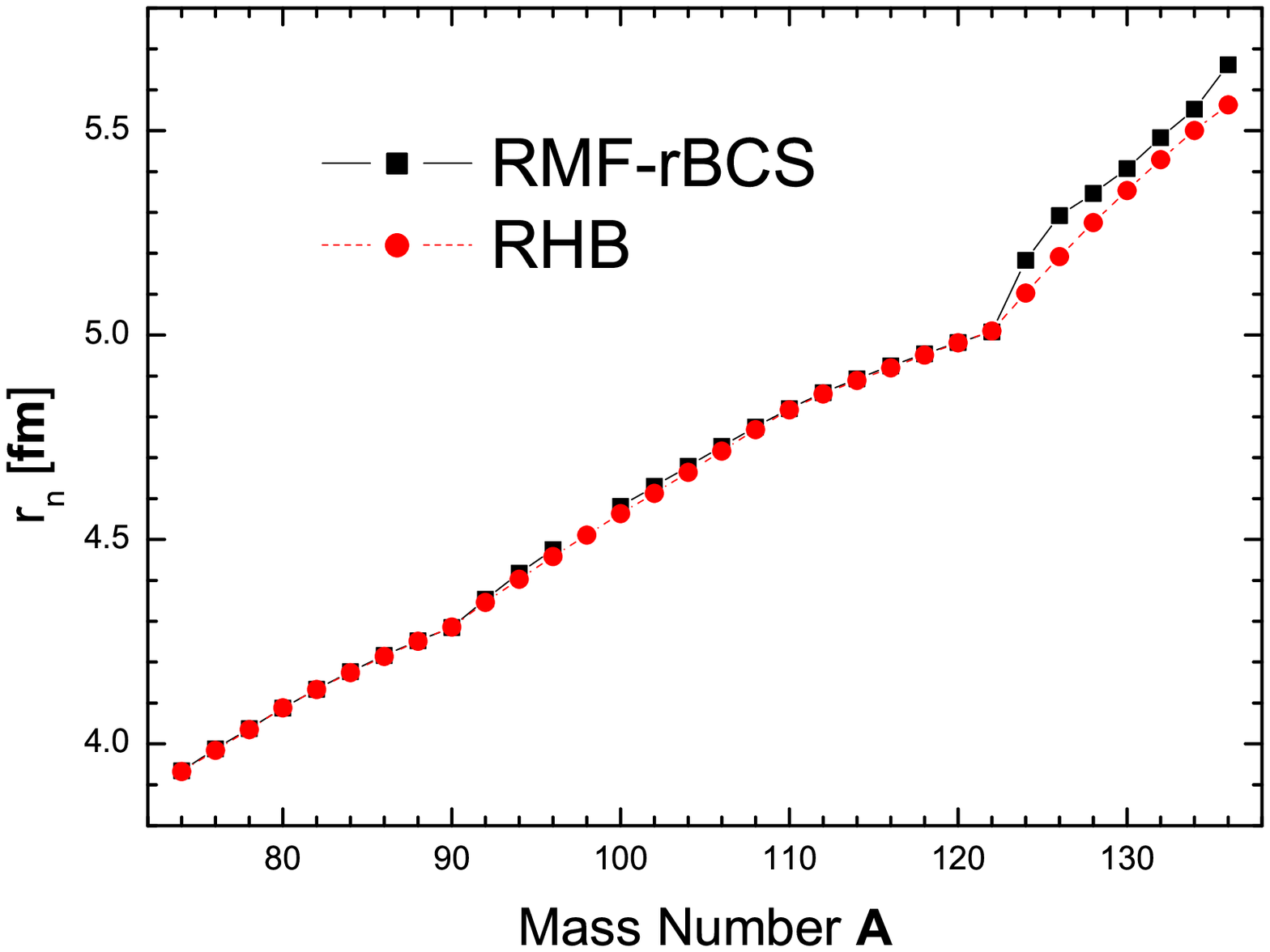}
\caption{\label{fig6.fig} Root mean square neutron radii of
even Zr isotopes as a function of the mass number A.}
\end{figure}

 The behaviour of the nuclear radii close to the drip line is very sensitive
 to the relative occupancy of the loosely bound states and the low-lying
 narrow resonances with high angular momenta. Thus if the diffusivity of the
 Fermi sea is increasing the pairs scatter from the loosely bound
 states to the narrow resonances, which are more localized around
 the nucleus. Consequently the nuclear radii might decrease if the
 average pairing gap is increasing.

  This effect can be also seen in the present calculations. In the RHB
 calculations the occupancy  of the narrow resonances with high angular
 momenta is larger than in RMF-rBCS calculations. Accordingly, as seen
 in Figure 6, the RHB radii are smaller than in RMF-BCS calculations.
 This situation is quite general since in the RHB or HFB calculations,
 based on a big energy cut off, the Fermi sea is usually more diffuse
 than in rBCS-type calculations, even if the pairing correlation
 energies might be rather similar in the two calculations.

\section{Conclusions}

In this paper we discussed how the resonant states can be treated
accurately in the RMF-BCS approach. The resonant states are
described through the scattering states located in the vicinity of
the resonance energies. These states are calculated by solving the
RMF equations with scattering-type boundary conditions for the
continuum spectrum. In the RMF-BCS equations the matrix elements
of the pairing interaction involving resonant states are
calculated by using the scattering states evaluated at the
resonance energies and normalised inside a finite region close to
the nucleus. The variation of the matrix elements of the pairing
interaction due to the widths of the resonant states is taken into
account by the derivative of the phase shift. This approximation
scheme, used previously in non-relativistic HF-BCS calculations,
is applied here for the neutron-rich Zr isotopes. It is shown that
the sudden increase of the neutron radii close to the neutron drip
line depends on a few resonant states close to the continuum
threshold. Including into the RMF-BCS calculations only these
resonant states one gets for the  neutron radii and neutron
separation energies practically the same results as in the more
involved RHB calculations.

\vskip 1.cm

{\bf Acknowledgments} \noindent We thank J. Meng for useful
discussions on RHB calculations. One of us (N.S) gratefully
acknowledges the COE Professorship Programme of Monkasho for
suporting his visit to Osaka University, where this work was
initiated, and the Swedish Programme for Cooperation in Research
and Higher Education (STINT).

\end{document}